\documentclass[prl,twocolumn,amsmath,amssymb,superscriptaddress,nofootinbib]{revtex4}

\usepackage{graphicx}
\usepackage{dcolumn}
\usepackage{bm}

\def\persec{{\rm s}^{-1}}

\def\Mpc{{\rm Mpc}}

\begin{document}

\title{Contraints on radiative dark-matter decay from the cosmic
  microwave background}

\author{Le Zhang}
\email{zl@cosmology.bao.ac.cn}
\affiliation{National Astronomical Observatories, Chinese Academy of
  Sciences, Beijing, 100012, China}
\affiliation{Department of Physics, Shandong University, Jinan,
  250100, China}

\author{Xuelei Chen}
\email{xuelei@cosmology.bao.ac.cn}
\affiliation{National Astronomical Observatories, Chinese Academy of
  Sciences, Beijing, 100012, China}

\author{Marc Kamionkowski}
\email{kamion@tapir.caltech.edu}
\affiliation{California Institute of Technology, Mail Code 130-33,
Pasadena, CA 91125, USA}

\author{Zong-guo Si}
\email{zgsi@sdu.edu.cn}
\affiliation{Department of Physics, Shandong University, Jinan,
  250100, China}

\author{Zheng Zheng}
\email{zhengz@ias.edu}
\thanks{Hubble Fellow}
\affiliation{Institute for Advanced Study, Einstein Drive, Princeton,
  NJ 08540, USA}

\date{April 18, 2007}

\begin{abstract}
If dark matter decays to electromagnetically-interacting
particles, it can inject energy into the baryonic gas and thus affect
the processes of recombination and reionization.  This leaves an
imprint on the cosmic microwave background (CMB): the large-scale
polarization is enhanced, and the small-scale temperature
fluctuation is damped.  We use the WMAP three-year data combined
with galaxy surveys to constrain radiatively decaying dark
matter. Our new limits to the dark-matter decay width are about
ten times stronger than previous limits. For dark-matter
lifetimes that exceed the age of the Universe, a limit of $\zeta
\Gamma_{\chi} < 1.7 \times 10^{-25}~\persec$ (95\% CL) is
derived, where $\zeta$ is the efficiency of converting decay
energy into ionization energy.  Limits for lifetimes short
compared with the age of the Universe are also derived.  We
forecast improvements expected from the Planck satellite.
\end{abstract}


\maketitle 

\section{Introduction}
After decades of research, the nature of the dark matter remains elusive.
Nevertheless, many properties of dark matter can be inferred from 
analysis of astrophysical processes that it might affect.
In particular, if the dark-matter (DM) particle is unstable, or if is
produced by decay of an unstable progenitor, then energy
released during the decay could affect recombination and
reionization.  This scenario received much  interest
\cite{CK04,D03} after the
release of the first-year Wilkinson Microwave Anisotropy Probe
(WMAP) results \cite{WMAP1}, which showed a large
temperature-polarization (TE) cross-correlation, indicating an
earlier epoch of reionization than is easily accommodated with
the $\Lambda$CDM model \cite{HH03}. In
this scenario, the decay of the DM particle 
supplies enough energy to provide early reionization.

Recently, the WMAP team has released data from three years of
observation \cite{WMAP3}, and the implications for
reionization are now more in line with the conventional
reionization scenario, though with a lowered matter density
$\Omega_m$ and power-spectrum normalization, the
demand on early and efficient star formation still exists
\cite{CF06}. While the original motivation of explaining
reionization with DM decay is now less compelling,
the calculations of the effects of DM decay on the CMB
can be turned now, with the new data, to deriving more stringent
constraints to decaying DM \cite{Kamionkowski:1999qc}.

Although 21-cm observations may some day be employed to
probe particle decay during the dark ages \cite{FOP06},
the best present constraints come from the CMB.
The CMB is affected by the ionization history in several ways: at large
scales, Thomson scattering by free electrons after reionization 
generates polarization from temperature anisotropies. 
This shows up as an enhanced TE cross-correlation and polarization
auto-correlation (EE).  This is why the new WMAP data, with greatly
improved measurements of large-angle polarization, will provide
much more stringent constraints to radiatively decaying DM.
At smaller scales, Thomson scattering
damps the primordial temperature anisotropy by a factor of
$e^{-2\tau}$. If the matter power spectrum can be measured
independently by other means, e.g. with large-scale-structure
(LSS) surveys, then this effect can also be used to constrain the
ionization history.

In the present work, we use the currently available CMB data,
including WMAP \cite{WMAP3}, ACBAR \cite{acb}, 
Boomerang \cite{Mo05}, CBI \cite{cbi}, and VSA \cite{vsa}, and the 
SDSS \cite{sdss} and 2dF \cite{2df05} galaxy surveys, to
constrain radiatively-decaying DM.
This work updates, expands, and improves upon our previous work
\cite{CK04} in several ways.  In addition to using the new WMAP
data, which is now far better suited for this analysis than
the first-year data,  we also include LSS data which, as we
shall see, provides some small additional
improvement.  We have also improved our methods, using now
Markov Chain Monte Carlo (MCMC) techniques for parameter
fitting, as we have done for analogous constraints to
DM annihilation in Ref.~\cite{Zh06}. This allows a
better understanding of the degeneracy between inferred
parameters, and it also provides more reliable error estimates.
This work also differs from Ref.~\cite{KK06} by
searching a broader range parameter space, doing a full
likelihood analysis, and including LSS.  In the regions where
our parameter spaces overlap, our limit is a factor of a few
better than theirs.  Refs.~\cite{MFP06} and \cite{Bean:2007rv} also
overlap with this work, but the former studies only several
particular DM candidates, rather than surveying the
entire decaying-DM parameter space, while the latter
considers heating/ionization only at recombination.

Our calculation of the effects of DM decay on
recombination, reionization, and heating of the intergalactic
medium, as well as their subsequent impact on the CMB
temperature and polarization power spectra follows that in
Ref.~\cite{CK04}.  The ionization rate from DM decay is
\begin{equation}
\label{eq:I}
     I_\chi = \chi_i f (f_\chi\Omega_\chi/\Omega_b) (
  m_b c^2/E_b)\Gamma_{\chi} e^{-\Gamma_{\chi} t},
\end{equation}
where $\Gamma_{\chi}$ is the decay width
(inverse lifetime) of the decaying particle, $m_b$ is the mass of the
baryon particle (hydrogen),
$(\Omega_\chi/\Omega_b)$ is the initial ratio of DM
mass to baryonic mass, $f$ is the fraction of the decay energy
deposited in the baryonic gas, $c$ is the speed of light,
$f_\chi=\Delta E/(m_\chi c^2)$, $\Delta E$ is the energy
released in the decay, $E_b=13.6$~eV is the binding energy
of hydrogen, and $t$ is the cosmological time.
The rate at which DM decay contributes to
the heating of the intergalactic medium (IGM) is
\begin{equation}
\label{eq:K}
K_\chi= \chi_h f (f_\chi\Omega_\chi/\Omega_b) m_b
  c^2\Gamma_{\chi} e^{-\Gamma_{\chi} t}.
\end{equation}
The decay energy deposited in the gas contributes both to the ionization and
heating of the gas. We model the division between these by an
ionizing fraction $\chi_i$ and heating fraction $\chi_h$
given by \cite{CK04} 
\begin{eqnarray}
     \chi_i = (1-x_e)/3, \qquad \chi_h = (1+ 2 x_e)/3,
\end{eqnarray}
where $x_e$ is the ionization fraction.  
The effects of decay on recombination can be 
determined completely by two independent parameters, 
$\Gamma_{\chi}$ and $\zeta$, where $\zeta = f f_\chi$, and is 
related to the quantity $\xi$ defined in Ref. \cite{CK04} 
by $\xi = \Gamma_{\chi} \zeta$.

We have modified the recombination code {\tt RECFAST}
\cite{SSS99}, which is used in
the CMB code {\tt CAMB} \cite{cambpaper} for the calculation of the
ionization history, to account for these extra contributions to 
ionization and heating. The {\tt CAMB} code is used by the MCMC code
{\tt COSMOMC} \cite{cosmomcpaper} as its CMB driver. We
also modified {\tt COSMOMC} so we could vary the new parameters
$\Gamma_\chi$ and $\zeta$ along with the cosmological parameters. For the
present study, we have considered the following set of 6 cosmological
parameters: $\{\Omega_b h^2, \Omega_c h^2, \theta, \tau, n_s, A_s\}$, $n_s$ and
$A_s$ are the the spectral index and amplitude of primordial
perturbations, and $\tau$ is the optical depth in the absence of
DM decay.  When we include LSS (which also encoded baryon acoustic
oscillation information) in our analysis, we also
marginalize over the galaxy bias $b$, and use data with $k<0.2 \Mpc/h
$ where the growth is linear.

When the ionization history is altered, the TE and EE spectra
are enhanced at large scales (small $\ell$). The TT spectrum at 
large $\ell$ is damped by a factor of $e^{-2\tau_{\rm tot}}$, where 
$\tau_{\rm tot}$ is the total optical depth.  
If the power-spectrum amplitude is not
known {\it a priori}, the large-$\ell$ suppression may appear as
a low-$\ell$ increase. However, if the matter power 
spectrum is obtained from galaxy surveys, the degeneracy
between $\tau_{\rm tot}$ and $A_s$ can be broken. Here we use the LSS
power spectrum obtained in the SDSS \cite{sdss} and 2dF
\cite{2df05} surveys.  We will see, though, that quantitatively
their inclusion improves the decaying-DM constraint
only a little, as the absolute amplitude is not known in this data
set, and the constraint comes only from the shape of the power spectrum.

\begin{figure} 
\includegraphics[width=0.43\textwidth]{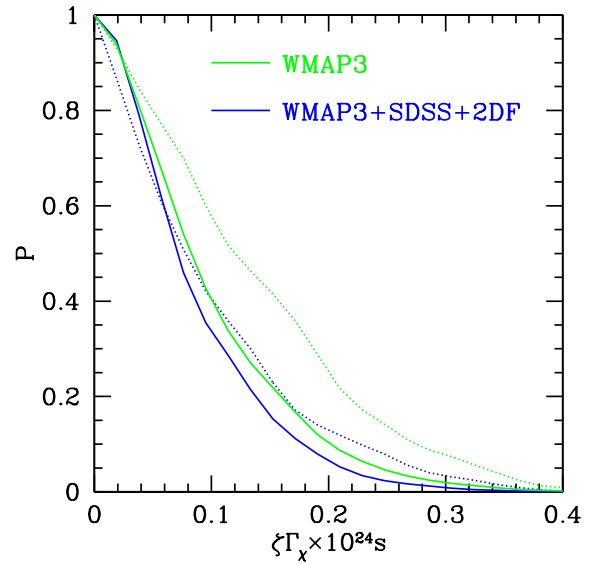}
\caption{\label{fig:long_1d} The marginalized probability 
distribution function (solid curve) and the relative mean likelihood
(dashed curve) of the $\zeta\Gamma_\chi$ 
parameter in the case of long lifetime,
for the CMB-only (WMAP 3 year) constraint (green upper
curves), and the CMB and LSS (SDSS+2dF) constraint (blue lower curves).  
The normalization is such that
the maximum of the function is 1.}
\end{figure}

If the decay lifetime is long compared with the age of the
Universe, the exponential factor in Eqs.~(\ref{eq:I}) and
(\ref{eq:K}) reduce to unity, so we can parameterize the effects
of DM decay fully by the combination $\zeta\Gamma_{\chi}$.
In Fig.~\ref{fig:long_1d}, we plot the marginalized probability
distribution function (PDF) of $\zeta\Gamma_{\chi}$ and the relative
mean likelihood, for both the constraint obtained from the CMB
only and also that obtained by including also LSS.
We find that these curves peak at $\zeta\Gamma_{\chi} =0$, indicating
that the CMB is consistent with no significant
DM decay.  The PDF drops rapidly at $\zeta\Gamma_{\chi} \sim
10^{-25}~\persec$, and the 95\% limit is at  
\begin{equation}
\label{eq:limitCMB}
\zeta\Gamma_\chi < 2.4 \times 10^{-25}~\persec,
\end{equation}
from the CMB only. The constraint is improved slightly with the 
addition of LSS data; the 95\% limit is then
\begin{equation}
\label{eq:limitLSS}
\zeta\Gamma_{\chi} < 1.7 \times 10^{-25}~\persec,
\end{equation}
in this case.  The degeneracy of the bias with the
galaxy-clustering amplitude limits the power of including LSS.
If the bias can be fixed by some other measurement, such as the
three-point correlation function, then there should be further
improvements better than those we have obtained here. For the present
data, if we assume $\zeta \leq 0.1$, we can exclude
models with lifetimes $\Gamma_{\chi}^{-1} < 5.9\times 10^{23}$~sec in
the long-lived case. This improvement over previous result \cite{CK04}
may provide strong constraint on new decaying-DM models such
as those given in Ref.~\cite{CFS07}

\begin{figure} 
\includegraphics[width=0.43\textwidth]{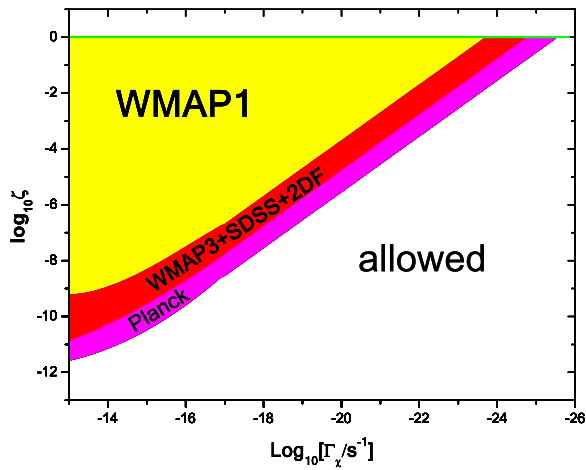}
\caption{\label{fig:ratio_gamma} The $2\sigma$ constraints on 
the decaying-DM parameter space.  We show the limits
from the WMAP1 analysis of Ref.~\protect\cite{CK04}, our new
constraint from WMAP3+SDSS+2dF, as well as our forecast for the
reach of Planck.}
\end{figure}

For a decay lifetime shorter than the age of the Universe at the epoch
of reionization ($\sim 10^{16} \sec$),
we need to fit the two parameters $\zeta$ and $\Gamma_\chi$
independently.  In Fig.~\ref{fig:ratio_gamma}, we plot the
excluded region ($2\sigma$) in the two-dimensional parameter
space ($\log_{10} \zeta$, $\log_{10} \Gamma_{\chi}$),
obtained after marginalization over the other parameters. The
region filled with red is excluded by the current CMB and LSS
data set. The boundary of this region is basically a
straight line with $\zeta \Gamma_\chi \sim$const. The line curves a little
bit at the short-lifetime end ($\Gamma_\chi \sim 10^{-13} \sec)$. 
A fit to the boundary curve is given by 
\begin{equation}
\label{eq:shortWMAP}
y=\left\{ \begin{array}{l}
                      6.77 + 3.96275x +0.25858x^2 + 0.00445x^3 \
                      \ (x > -17), \\ 
                        -24.75-x  \ \ \ (x < -17),
                        \end{array}
                        \right. 
\end{equation}
where $x= \log_{10}{\Gamma_\chi}$, and $y= \log_{10}{\zeta}$.  
We have restricted the parameter
space to $\Gamma_\chi < 10^{-13}~\persec$, because for still shorter
lifetime, much of the decay would happen before recombination
and would thus not affect the ionization history of the Universe. 

We will soon have CMB data of still better quality from the
Planck satellite \cite{Planck}, to be launched
soon.  Anticipating this development, we use the Fisher-matrix
formalism \cite{fisher,Zh06} to forecast the future measurement
error. We assume a sky coverage of $0.65$ and adopt a
fiducial model that best fits the WMAP three-year data. Because Planck
will determine the B-mode polarization, we evaluate the
constraint derived from the TT, TE, EE, and BB spectra, and use the
covariance matrices of Ref.~\cite{kks}.  The resulting
constraint is also plotted in Fig.~\ref{fig:ratio_gamma}, and a fit to
it is given by
\begin{equation}
\label{eq:Planck}
y=\left\{ \begin{array}{ll}
3.6 + 2.645x +0.114x^2 \ \ (x > -17), \\
-25.531-x  \ \ \ (x < -17),
\end{array}\right. 
\end{equation} 
where $x= \log_{10}{\Gamma_\chi}$, and $y= \log_{10}{\zeta}$.

\begin{figure} 
\includegraphics[width=0.41\textwidth]{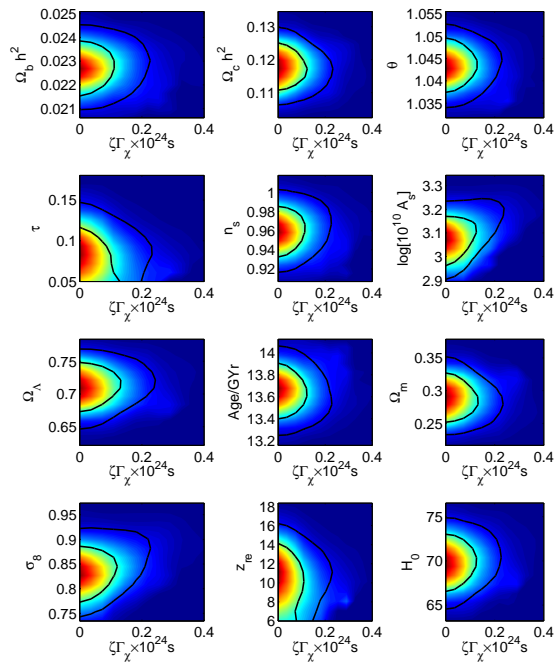}
\caption{\label{fig:long_2D} The 2D contours of
the distribution of $\zeta\Gamma_\chi$ and the 
background parameters for WMAP3+SDSS+2dF data. The color is for
the relative mean likelihood}
\end{figure}

Dark-matter decay may also affect the global fitting of 
cosmological parameters. We expect the decay parameter 
$\zeta\Gamma_\chi$ to correlate with optical depth $\tau$, as
both of these are related to the ionization history.  We also expect it to 
correlate with $A_s$ and $n_s$, as  scattering from electrons damps
small-scale CMB anisotropy. There should also be some correlation
with $\Omega_c$, as this fixes the rate of input of decay
energy.  We
plot the 2D contours of the $\zeta\Gamma_\chi$ parameter with 
other cosmological parameters in Fig.~\ref{fig:long_2D}. There is
indeed some correlation with the above parameters. By contrast,
if we consider additional ionization and heating due to
DM annihilation \cite{Zh06} (rather than decay), the
correlation is very weak because DM annihilation occurs
mainly during the earlier epoch of recombination when the
DM density is higher \cite{Zh06}.  However, the
correlation is perhaps not quite as strong as one might have
anticipated from the Fisher-matrix analysis \cite{CK04}. This,
we believe, is because the plotted contours follow
marginalization, during which the correlation is weakened
somewhat.

We have studied the impact of decaying DM
on the ionization history and its subsequent effects on the
CMB. Our investigation is largely 
model-independent; i.e., it does not depend on the particular type of
particle, but only on the decay width $\Gamma_\chi$ and
energy-conversion efficiency $\zeta$. Dark-matter decay may
affect the temperature and ionization fraction of the baryonic
gas. It may also help produce or destroy molecular
hydrogen, which plays an important role in the formation of the first
stars \cite{RMF06}. We have not considered such effects in
this work, but treated DM decay and star
formation as two independent processes. We obtained constraints on
DM decay by using data from WMAP's three-year results
and the the SDSS and 2dF galaxy surveys.  We have derived limits
on decay parameters [cf.,~Eqs.~(\ref{eq:limitCMB})--(\ref{eq:shortWMAP})].
The constraint on decaying DM will be further improved
with future CMB experiments such as the Planck satellite
[cf.,~Eq.~(\ref{eq:Planck})], and/or direct observation of the
dark ages with 21-cm surveys \cite{FOP06}.

\bigskip
Our MCMC chain
computation was performed on the Supercomputing Center of 
the Chinese Academy of Sciences and the Shanghai Supercomputing
Center. This work is supported by
the National Science Foundation of China under the Distinguished Young
Scholar Grant 10525314, the Key Project Grant 10533010, by the 973
Program No. 2007CB815401, and by the
Chinese Academy of Sciences under grant KJCX3-SYW-N2.  MK was
supported by DoE DE-FG03-92-ER40701, NASA NNG05GF69G, and the
Gordon and Betty Moore Foundation. ZZ was supported by NASA through
Hubble Fellowship HF-01181.01-A awarded by the Space Telescope Science
Institute, which is operated by the Association of Universities for
Research in Astronomy, Inc., for NASA under contract NAS 5-26555.

\newcommand\AAP[3]{Astron. Astrophys.{\bf ~#1}, #2~ (#3)}
\newcommand\AL[3]{Astron. Lett.{\bf ~#1}, #2~ (#3)}
\newcommand\AP[3]{Astropart. Phys.{\bf ~#1}, #2~ (#3)}
\newcommand\AJ[3]{Astron. J.{\bf ~#1}, #2~(#3)}
\newcommand\APJ[3]{Astrophys. J.{\bf ~#1}, #2~ (#3)}
\newcommand\APJL[3]{Astrophys. J. Lett. {\bf ~#1}, L#2~(#3)}
\newcommand\APJS[3]{Astrophys. J. Suppl. Ser.{\bf ~#1}, #2~(#3)}
\newcommand\MNRAS[3]{Mon. Not. R. Astron. Soc.{\bf ~#1}, #2~(#3)}
\newcommand\MNRASL[3]{Mon. Not. R. Astron. Soc.{\bf ~#1}, L#2~(#3)}
\newcommand\NPB[3]{Nucl. Phys. B{\bf ~#1}, #2~(#3)}
\newcommand\PLB[3]{Phys. Lett. B{\bf ~#1}, #2~(#3)}
\newcommand\PRL[3]{Phys. Rev. Lett.{\bf ~#1}, #2~(#3)}
\newcommand\PR[3]{Phys. Rep.{\bf ~#1}, #2~(#3)}
\newcommand\PRD[3]{Phys. Rev. D{\bf ~#1}, #2~(#3)}
\newcommand\SJNP[3]{Sov. J. Nucl. Phys.{\bf ~#1}, #2~(#3)}
\newcommand\ZPC[3]{Z. Phys. C{\bf ~#1}, #2~(#3)}
\newcommand\SCI[3]{Sci.{\bf ~#1}, #2~(#3)}

\end{document}